\documentclass{sig-alternate}

\usepackage{amsfonts}
\usepackage{amssymb}
\usepackage[utf8]{inputenc}
\usepackage{times}
\usepackage[english]{babel}
\usepackage{graphicx}
\usepackage{mathtools}
\usepackage{bm}
\usepackage{comment}
\usepackage{epstopdf}
\usepackage{stfloats}
\usepackage{cite}
\usepackage{dsfont}
\usepackage{url}
\usepackage{amsmath}
\usepackage{psfrag}
\usepackage{subfigure}
\usepackage{multirow}

\begin{document}
%
\conferenceinfo{WNS3}{May~7 2014, Atlanta, Georgia, USA}

\title{Extending the Energy Framework for\\ 
Network Simulator 3 (ns-3)}

\numberofauthors{1} 
%
\author{Cristiano Tapparello, Hoda Ayatollahi, Wendi Heinzelman\\[1.5mm]
       \affaddr{Department of Electrical and Computer Engineering, University of Rochester, Rochester, NY, USA}
       }

\maketitle

In the last few years, the capability of Network Simulator 3 (ns-3) of simulating different aspects of wireless networks has increased rapidly, such that it now provides a wide range of models of real world objects, protocols and devices~\cite{ns3}. 
Simulation of communication systems and network protocols over realistic device operations is seen as a necessary task before implementation, because it allows for a flexible and fast, but still accurate, testing of the system evolution. Additionally, efficiently managing the energy consumptions of the different elements is a major requirement for an efficient design of wireless networks, since many wireless devices are battery operated.

In this regard, the authors in~\cite{wu2011} presented an ns-3 energy framework that allows users to simulate the energy consumption at a node as well as to determine the overall network lifetime under specific conditions. This framework adds sufficient support to ns-3 to devise simulations that include the energy consumption of the communication network. In order to do this, this framework defines the concept of an {\it Energy Source}, which represents an abstraction of the way in which the node is powered, a {\it Device Energy Model}, that defines models for the energy consumption of the different elements that compose the node, and several methods that provide different types of energy information (e.g., residual energy, current load, etc.) to other ns-3 objects external to the framework. The framework is developed with the objective of allowing the interoperability of different energy source and device energy consumption models, and it allows easy integration of new models. Moreover, some implementations of the energy source and device energy models are provided, so that an ns-3 user is able to incorporate them into their existing simulations. 

The increasing demand for battery operated devices with longer lifetimes has required researchers to explore energy availability from a different perspective, starting from the hardware itself. While the battery technology keeps improving, with the recent advancement in real life wireless devices, devices that are able to harvest energy from the environment, e.g., in the form of solar, thermal, vibrational or radio energy, are now commercially available.

Given the above, the problem of designing optimal transmission protocols for energy harvesting wireless networks has recently received considerable attention~\cite{Sudevalayam11,Ozel11,Sharma10}. When using an energy harvesting source, the objectives of these protocols are fundamentally different than those of using a traditional energy source: rather than focusing on minimizing the maximum energy and adapting the operations according to the residual energy, algorithms needs to shift the optimization to the maximum rate at which the energy can be used~\cite{Kansal07}. The need for an accurate modeling of the energy harvesting process, and a consequent redesign of the simulation framework to include it, is thus fundamental.

While the current ns-3 energy framework (version 3.19) allows the definition of new energy sources that incorporate the contribution of an energy harvester, the integration of an energy harvester component into an existing energy source, as well as the possibility of evaluating the interaction between different energy sources and harvester models, is not straightforward using the existing energy framework.

\begin{figure}
 \centering  \includegraphics[width=0.8\columnwidth]{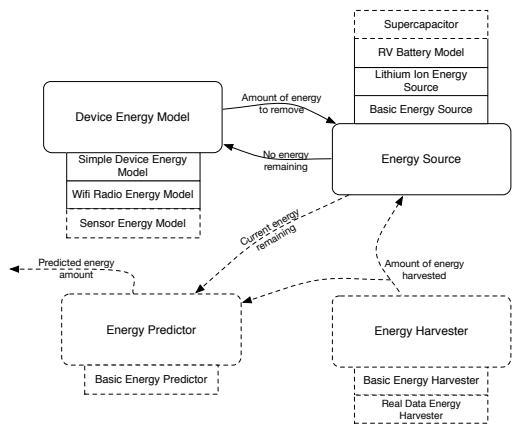}
\caption{Energy Framework. Our contribution is shown by the dashed lines.}
\label{fig:energy_framework}
\end{figure}

To this end, in this poster we propose an extension of the energy framework currently released with ns-3 in order to explicitly introduce the concept of an energy harvester. By doing so, different energy harvester models can be developed as independent ns-3 objects that can be connected to the current and future energy source implementations. Starting from the definition of the general interface, we then provide the implementation of two simple models for the energy harvester: 1) a basic energy harvester, that provides a time-varying, uniformly distributed amount of power, and 2) an energy harvester that recharges the energy source with an amount of power gathered from a dataset of real solar panel measurements~\cite{dataset}. 

Our contribution to the energy framework extends beyond the introduction and example implementations of the energy harvester. In particular, we extended the set of implementations of the current energy framework to include a model for a supercapacitor energy source and a device energy model for the energy consumption of a sensor. Moreover, we introduced the concept of an {\it energy predictor}, and we provide an example implementation based on the energy prediction model described in~\cite{Cammarano12}. As the name suggests, the idea behind the energy predictor is to gather information from the energy source and harvester and use this information to predict the amount of energy that will be available in the future. The information provided by this module can then be used to develop energy efficient protocols that can capitalize not only on the knowledge of the energy availability at a given time but also on a forecast of its future availability.

\begin{figure}
 \centering  \includegraphics[width=0.8\columnwidth]{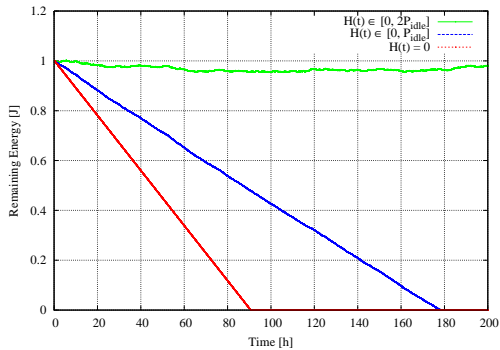}
\caption{Remaining energy as a function of time for a simple node that senses the environment every minute, for different values of the energy harvested $H(t)$. Operational values: $P_{idle} = 3 \mu W$, $P_{active} = 15 \mu W$. $H(t)$ is uniformly distributed in the relative interval, and its value is updated every $5$ minutes.}
\label{fig:example_energy}
\end{figure}
\begin{figure}
 \centering  \includegraphics[width=0.8\columnwidth]{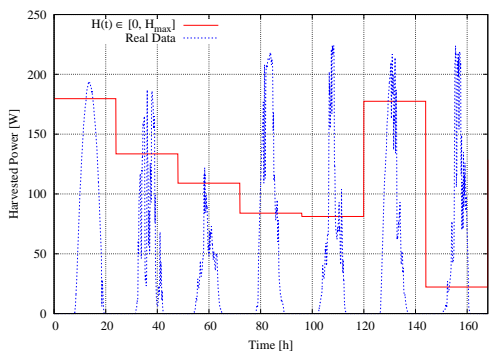}
\caption{Harvested power as a function of time, for the basic energy harvester and the dataset of real solar panel measurements~\cite{dataset}. $H(t)$ is uniformly distributed in $[0, H_{max}]$, with $H_{max} = 220 W$, and its value is updated every $24$ hours.}
\label{fig:harvester_comparison}
\end{figure}

\begin{figure}
 \centering  \includegraphics[width=0.8\columnwidth]{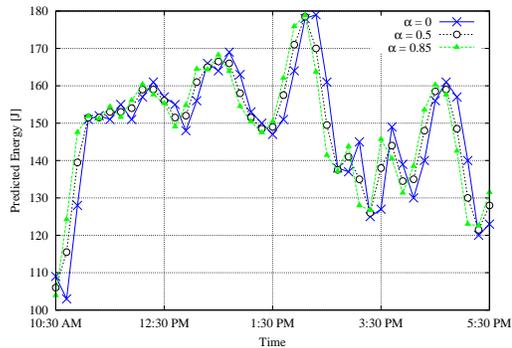}
\caption{Energy predicted by the basic energy predictor module, for different values of $\alpha$.}
\label{fig:energy_predictor}
\end{figure}

A diagram of the elements that compose the extended energy framework is presented in Figure~\ref{fig:energy_framework}. In order to provide an example of the results that can be obtained with our extension of the ns-3 energy framework, we consider a simple node powered by a {\it Basic Energy Source} and a {\it Basic Energy Harvester}, that performs periodic sensing of the environment through a basic {\it Sensor} implementation. In Figure~\ref{fig:example_energy}, we plot the remaining energy of the node as a function of time, for different values of harvested energy. As expected, the presence of the energy harvester will extend the lifetime of the node for a time proportional to the amount of harvested energy. In Figure~\ref{fig:harvester_comparison}, we compare the power provided by the basic energy harvester with the harvester based on the real dataset. Finally, in Figure~\ref{fig:energy_predictor} we plot the output of the basic energy predictor module. According to the model presented in~\cite{Cammarano12}, the predicted energy $\hat{E}_{t+1}$ is computed as $\hat{E}_{t+1} = \alpha C_{t} + (1 - \alpha) E^{d}_{t+1}$, where $C_{t}$ represents the energy harvested during timeslot $t$ of the current day, $E^{d}_{t+1}$ is the energy harvested during timeslot $t + 1$ of a stored day $d$, and $0 \leq \alpha \leq 1$ is a weighting parameter.

As future work, we plan to further extend the energy framework to include several implementations of the different elements that compose the framework, as well as to provide an application that links them together. In this regard, we are currently working on the implementation of an application that simulates a distributed source coding scenario for a wireless network powered via energy harvesting. 

As a result of these efforts, we believe that our contributions to the ns-3 energy framework will provide a useful tool to enhance the quality of simulations of energy-aware wireless networks.

\bibliographystyle{IEEEtran}
\bibliography{IEEEabrv,biblio}

\end{document}